\documentclass[11pt,letterpaper]{article}
\usepackage{cogsys}
\usepackage[T1]{fontenc}
\usepackage{times}
\usepackage[pdftex]{graphicx} % use this when importing PDF files
\usepackage{dialogue}
\usepackage{longtable}
\usepackage{tikz}
\usepackage{wrapfig}

{\obeyspaces\gdef {\ }}
\global\newbox\codebox
\global\newbox\savedcodebox
\gdef\sverbatim{\bgroup\def\endsverbatim{\egroup\egroup\egroup\mbox{\box\codebo\
x}}\def\savecode{\egroup\egroup\egroup\global\setbox\savedcodebox\copy\codebox}\
\def\par{\egroup\vspace{-0.3em}\hbox\bgroup}\tt\obeylines\obeyspaces\global\set\
box\codebox\vbox\bgroup\hbox\bgroup}

\newcommand{\avmplus}[1]{{\setlength{\arraycolsep}{0.8mm}
                       \renewcommand{\arraystretch}{1.2}
                       \left[
                       \begin{array}{l}
                       \\[-2mm] #1 \\[-2mm] \\
                       \end{array}
                       \right]
                    }}
\newcommand{\att}[1]{{\mbox{\small {\bf #1}}}}
\newcommand{\attval}[2]{{\mbox{\small {\sc #1}}\ =\ {{#2}}}}

\newcommand{\attvaltyp}[2]{{\mbox{\small{\sc #1}}\ =\ {\myvaluebold{#2}}}}
\newcommand{\attvaltypnoatt}[2]{{\mbox{$\;\;\;\;\;\;\;\;\;\;$} \  \ {\myvaluebold{#2}}}}

\newcommand{\myvaluebold}[1]{{\mbox{\small {\bf #1}}}}

\usepackage{natbib}
\cogsysheading{X}{20XX}{1-6}{X/20XX}{X/20XX}
\ShortHeadings{Multimodal Continuations for HRI}
              {N.\ Krishnaswamy and J.\ Pustejovsky}

\begin{document} 

\title{Multimodal Continuation-style Architectures for Human-Robot Interaction}
 
\author{Nikhil Krishnaswamy}{nkrishna@brandeis.edu}
\author{James Pustejovsky}{jamesp@brandeis.edu}
\address{Department of Computer Science, Brandeis University, 
         Waltham, MA 02148 USA}
\vskip 0.2in
 
\begin{abstract}
We present an architecture for integrating real-time, multimodal input into a computational agent's contextual model.  Using a human-avatar interaction in a virtual world, we treat aligned gesture and speech as an {\it ensemble} where content may be communicated by either modality.  With a modified nondeterministic pushdown automaton architecture, the computer system: (1) consumes input incrementally using continuation-passing style until it achieves sufficient understanding the user's aim; (2) constructs and asks questions where necessary using established contextual information; and (3) maintains track of prior discourse items using multimodal cues.  This type of architecture supports special cases of pushdown and finite state automata as well as integrating outputs from machine learning models. We present examples of this architecture's use in multimodal one-shot learning interactions of novel gestures and live action composition.
\end{abstract}

\section{Introduction} 
\label{sec:intro}
Unlike interaction with other types of interactive agents (e.g., chatbots or personal digital assistants), human-robot interaction inherently requires multi-modality.  Robotic agents are {\it embodied} and {\it situated} which affords robots the ability to affect the real world, but also requires them to have accurate and robust interpretive capabilities for multiple {\it input modalities}, which must run in real time.  In addition, a robot must be able to communicate with its human interlocutors using all {\it communicative modalities} humans may use, including natural language, body language, gesture, demonstrated action, emotional cues, etc.  As robots take on more human-like appearances, this becomes even more important, as there exists a gulf between expectations that the robot will communicate and understand things in a human-like way and its actual multimodal capability \citep{luger2016like}. 

Computers as collaborators (a la social robotics) require architectures that enable communication with humans naturalistically and multimodally, as humans do with each other.  These architectures must be able to capture not only natural language (spoken, written, sign, etc.), but also gesture and body language, dynamic discourse semantics \citep{asher1998common}, affect and emotion \citep{scheutz2006utility}, etc., all in context.  Context, in the case of a human-robot interaction, comes in large part from the relative embodiment of the human and the robot or agent, and their situatedness with respect both to the scene that they both inhabit and to each other (``{\it co-situatedness}" \citep{pustejovsky2017creating}).  But situatedness in the virtual world or the agent's internal model alone is not enough: the agent must be able to situate and model itself in the physical world of its interlocutors and interpret contextualized input relative to that space \citep{pustejovsky2019situational}.  This in turn entails that the agent be {\it socially situated} and {\it socially embedded}, following \cite{dautenhahn2002embodied} and a multimodal interface is at minimum a {\it social interface}, following \cite{breazeal2003toward}.

A number of architectures exist that handle at least some of the technical requirements of these kinds of social robots or agents, from natural language processing to motor control (\cite{jaimes2007multimodal} and \cite{goodrich2008human} provide surveys of HRI and HCI, including architectures, as of time of publication).  We present a computational framework for capturing and reasoning over conversational and situational context, and driving the agent's actions or responses in the world.  A number of different particular reasoning architectures can be created using this framework, which is built on top of the VoxML modeling language \citep{PUSTEJOVSKY16.1101} as its representation of object and event semantics.  It exploits continuation-passing style \citep{van1966recursive,reynolds1993discoveries,van2010computational} to retrieve situational contextual information and compose it with object and event properties to conduct reasoning at runtime, and is designed modularly to facilitate integration with other robotic architectures, such as {\sc DIARC} \citep{scheutz2007first}, POMDP approaches \citep{zhang2017robot}, or reinforcement learning \citep{peters2008reinforcement}.

Since the architectures described here are currently deployed on systems using an agent in a virtual world, but are being developed for use in interactions with physical robot, we will use the term {\it human-avatar interaction} (HAI) for interaction between a human and an {\it embodied, situated agent}, be it an animated avatar in a virtual world or a robotic agent in the physical world.

\iffalse
\section{Related Work}
\label{sec:related}

Possibly integrate into intro

Need cites for other HRI architecture (besides DIARC)--e.g., POMDPs, see papers from 2017 AAAI SSS--continuation foundations, possible existing combos
\fi

\section{Interactive and Formal Structure}
\label{sec:structure}

Assume a task-oriented HAI dialogue reproducing most conventions of human-to-human task-oriented dialogue (e.g., cooperation, responsiveness, disambiguation, etc.).  Interlocutors might refer to objects and actions in any order, i.e., a single utterance: ``put the knife in the blue cup"; or a multi-step dialogue specifying entities and actions involved.  Formally, this requires that arguments be applied to previously cached predicates and vice versa at runtime, and in turn requires a representation capable of executing methods where arguments can be raised to the type required.

\subsection{Interactive Structure}
\label{ssec:interact}
Implementationally, the system in which we deploy our multimodal architectures is build on the VoxSim software \citep{krishnaswamy2016multimodal}, itself build on the VoxML platform and modeling language.  VoxSim consumes input from real-time gesture and speech recognition clients, as described in \cite{krishnaswamy2017communicating} and \cite{narayana2018cooperating}, and conducts real-time inference, reasoning, and disambiguation using a visualized and situated simulated environment as the computer's representation of the shared world between agent and human.

\begin{figure}[h!]
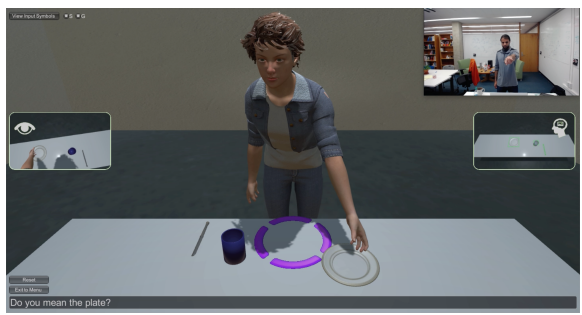

    \centering
    \begin{tabular}{|l|l|}
    \hline
    \begin{minipage}{.5\textwidth}
    \begin{dialogue}
    \speak{human}  ``The plate." [A]
    \speak{agent} \direct{\refer{agent} reaches for plate} 
     ``Okay, go on."
    \speak{human}  ``Put it in front of you." [B]
    \speak{agent} \direct{\refer{agent} puts plate in front of itself} ``Okay."
    \end{dialogue}
    \end{minipage}
& 
    \begin{minipage}{.5\textwidth}
    \begin{dialogue}
    \speak{human}  ``The plate." [A]
    \speak{agent} \direct{\refer{agent} reaches for plate} ``Okay, go on."
    \speak{human} \direct{\refer{human} points} ``Put it there." [B]
    \speak{agent} \direct{\refer{agent} puts plate at indicated location} ``Okay."
    \end{dialogue}
    \end{minipage}
    \\
\hline
    \end{tabular}
    \caption{Dialogues---using only language (L) and language with gesture (R)}
    \label{fig:dialogues}
\end{figure}

Within this environment, consider two dialogues---one conducted entirely with language and one with a combination of language and gesture (Fig.~\ref{fig:dialogues}). Dialogue~\ref{fig:dialogues}L is conducted using only language input, and in this example language suffices to give the directions necessary.  However, in many cases, HAI may require the ability to indicate information using a different method.  For instance, direct grounding to a location may be needed because the location is too complicated to describe in language, or simply for efficiency:

\begin{figure}[h!]
    \centering
\begin{tabular}{ll}
    \begin{minipage}{.5\textwidth}
    \includegraphics[width=\textwidth]{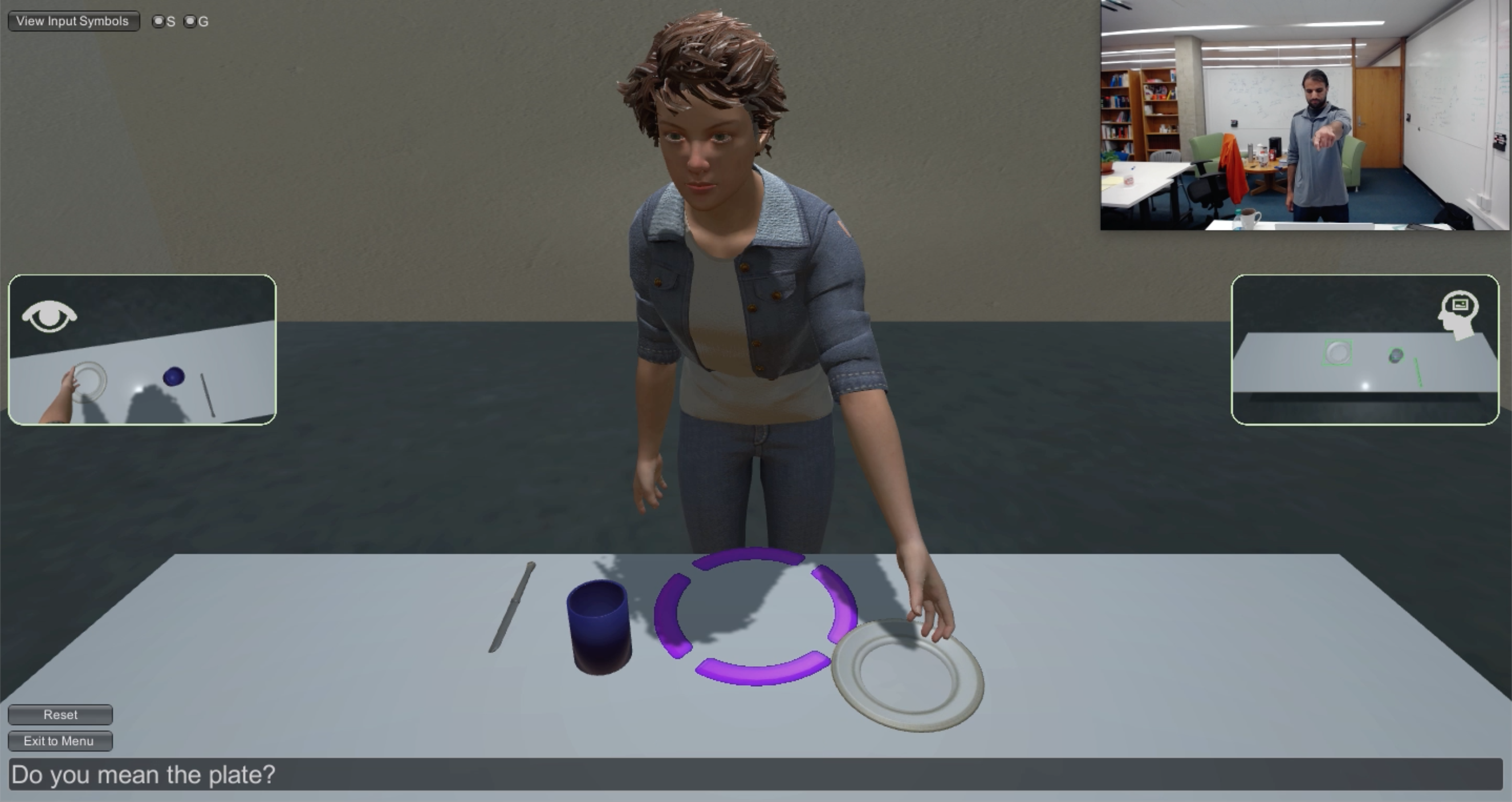}
\end{minipage}
&
\begin{minipage}{.5\textwidth}
{\tiny
\def\baselinestretch{1.1}
$\avmplus{\att{point}\\
	%\attval{lex}{...}\\
	\attval{type}{\avmplus{
		\attvaltyp{head}{assignment}\\
		\attval{args}{\avmplus{
			\attvaltyp{a$_1$}{$\mbox{x:agent}$}\\
			\attvaltyp{a$_2$}{$\mbox{y:finger}$}\\
			\attvaltyp{a$_3$}{$\mbox{z:location}$}\\
			\attvaltyp{a$_4$}{$\mbox{w:physobj[]$\bullet$}$}\\
			\attvaltypnoatt{\;}{$\mbox{location}$}
		}}\\
		\attval{body}{\avmplus{
			\attvaltyp{e$_1$}{$extend(x,y)$}\\
			\attvaltyp{e$_2$}{$def(vec($}\\
			\attvaltypnoatt{\;}{$x \rightarrow y \times z),$}\\
			\attvaltypnoatt{\;}{$as(w))$}
		}}
	}}
}$
\def\baselinestretch{1.9}}
    \end{minipage}
\end{tabular}
\caption{\label{fig:gesturepoint}L: Example multimodal interaction accompanying Dialogue~\ref{fig:dialogues}R.  Upper-right inset shows human pointing to location; purple target in main image shows interpreted location of deixis in virtual world.  R: VoxML typing of [[{\sc point}]] \citep{PUSTEJOVSKY16.1101}.  {\sc e$_{2}$} defines the target of deixis as the intersection of the vector extended in {\sc e$_{1}$} with a location, and reifies that point as a variable $w$. {\sc a$_{4}$}, shows the compound binding of $w$ to the indicated region and objects within that region \cite{ballard1997deictic}.}
\end{figure}

In both dialogues, the human specifies the object to be manipulated in [A] and then specifies a location in [B] along with an action.  In Dialogue~\ref{fig:dialogues}R, the human's use of a demonstrative word, ``there" is accompanied by a deictic gesture (Fig.~\ref{fig:gesturepoint}), which grounds the demonstrative to a specific location or objects at that location.  ``There" selects for the location indicated by deixis, not an object.

\begin{wrapfigure}{l}{.4\textwidth}
    \includegraphics[width=.38\textwidth]{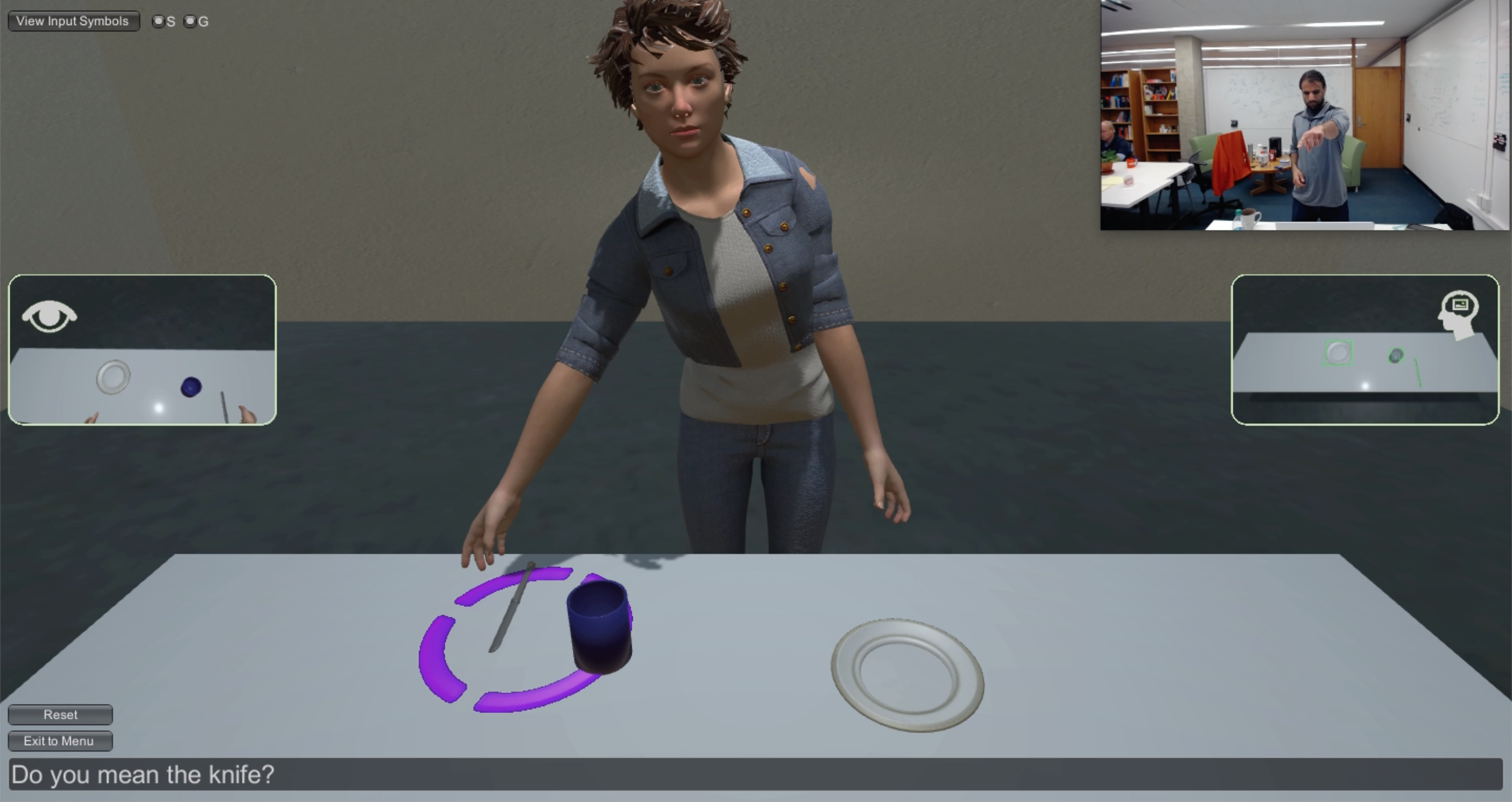}
    \vspace{-4mm}
    \caption{\label{fig:gesturedemo2}Deixis to region with objects.}
    \vspace{-10mm}
\end{wrapfigure}

The human can indicate object properties with accompanying deixis to single out objects in a region subject to particular properties.  Thus given a deictic gesture as in Fig.~\ref{fig:gesturedemo2}, ``cup," ``that cup", ``the blue cup," or even ``in that blue cup" or ``put the knife in the blue cup" single out the same object in the region, specifying objects, locations, or actions using multiple modalities.

\subsection{Formal Structure}
\label{ssec:formal}

Formally, we can define the vocabulary of the interaction, consisting of sequences of ``moves" by both interlocutors (as defined in \cite{krishnaswamy2018LREC,pustejovsky2018actions}), using the format of a context-free grammar (CFG) where the nonterminals represent sets of particular input symbols and the terminals represent the particular content or intended response communicated by those input symbols.
%For instance, we might define $O \rightarrow \delta D$ as part of our grammar, where $O$ is a sequence of turns used to indicate an object, $\delta$ is a deictic gesture, and $D$ is set of moves in which the human assists the agent in disambiguating a reference to a particular object. 
We can define a portion of the interactive ``grammar" as follows, where unexpanded nonterminals represent parsed sentences or phrases or variations on individual gestures:

\begin{table}[h!]
    \centering
    \begin{tabular}{lll}
        {\bf Grammar} & {\bf Legend} \\
        $S \rightarrow OA|AO$ & $O$: define object & $\omega$: static iconic gesture (object) \\
        $O \rightarrow \delta|\delta D|\omega|\omega D|N|ND$ & $A$: define action & $\alpha$: dynamic iconic gesture (action) \\
        $A \rightarrow \alpha|\alpha D|V|VD|P|PD$ & $D$: disambiguate & $\delta$: deictic gesture \\
        $D \rightarrow \delta|\delta D|P|PD|N|ND|y|yD|n|nD$ & $V$: verb phrase & $y$: affirmative response \\
        & $N$: noun phrase & $n$: negative response \\
        & $P$: prep. phrase & \\
    \end{tabular}
    \vspace{-4mm}
    \caption{Interactive grammar snippet}
    \label{tab:grammar}
\end{table}
\vspace{-2mm}

\iffalse
\begin{center}
\begin{longtable}{lll}
        {\bf Grammar} & {\bf Legend} \\
        $S \rightarrow OA|AO$ & $O$: define object & $\omega$: static iconic gesture (object) \\
        $O \rightarrow \delta|\delta D|\omega|\omega D|N|ND$ & $A$: define action & $\alpha$: dynamic iconic gesture (action) \\
        $A \rightarrow \alpha|\alpha D|V|VD|P|PD$ & $D$: disambiguate & $\delta$: deictic gesture \\
        $D \rightarrow \delta|\delta D|P|PD|N|ND|y|yD|n|nD$ & $V$: verb phrase & $y$: affirmative response \\
        & $N$: noun phrase & $n$: negative response \\
        & $P$: prep. phrase & \\
    \caption{Interactive grammar snippet}
    \label{tab:grammar}
\end{longtable}
\end{center}
\fi
At each step, the disambiguation symbol $D$ represents the acquisition of information the agent still needs to know in order to complete some action initiated or requested by the human.  Furthermore, the order of instructions may vary, requiring that the agent be able to hold previously acquired information ``in reserve" pending further instruction or answers to disambiguatory questions.

Due to the large number of terminal symbols in the ``grammar" in even a superficial system, creating new states for every possible contextual configuration is computationally inefficient.  For instance, if there are three objects in the scene, an object disambiguation sequence should not have to proceed through three distinct states, waiting for a {\it yes} or {\it no} each time, to confirm which of the three objects should be the focus, when instead it can recurse through the same state with a different argument until {\it yes} is received, and then proceed to the handling of the argument, or proceed to another state when all possible arguments are exhausted. 

\iffalse
\begin{center}
\begin{tikzpicture}[->,auto,node distance=3cm,
  thick,main node/.style={circle,draw}]

  \node[main node] (1) {{\it a}};
  \node[main node] (2) [right of=1] {{\it b}};
  \node[main node] (3) [right of=2] {{\it c}};
  \node[main node] (4) [right of=3] {{\it d}};

  \path[every node]
    (1) edge node [above] {label} (2)
    (2) edge node [right] {} (3)
    (3) edge[bend right] node [below] {label} (4)
    (4) edge[loop above] node [left] {} (4);
\end{tikzpicture}
\end{center}
\fi

\begin{figure}[h!]
    \centering
    \begin{tabular}{|c|c|}
    \hline
	\begin{tikzpicture}[->,auto,node distance=2cm,
  	thick,main node/.style={circle,draw},phantom node/.style={circle,draw}]

  	\node[main node] (1) {{\it ?knife}};
  	\node[main node] (2) [right of=1] {{\it ?cup}};
  	\node[main node] (3) [right of=2] {{\it ?plate}};
  	\node[phantom node] (4) [right of=3,below=3] {};
 	 \node[phantom node] (5) [right of=3,above=3] {};

 	 \path[every node]
    	(1) edge[bend right] node [below] {no} (2)
    	(1) edge[bend left] node [above] {yes} (5)
   	 (2) edge[bend right] node [below] {no} (3)
    	(2) edge[bend left] node [above] {yes} (5)
    	(3) edge[bend right] node [below] {no} (4)
    	(3) edge[bend left] node [above] {yes} (5);
	\end{tikzpicture}
         &
	\begin{tikzpicture}[->,auto,node distance=2cm,
  	thick,main node/.style={circle,draw},phantom node/.style={circle,draw}]

  	\node[main node] (1) {{\it ?arg}};
  	\node[phantom node] (2) [right of=1,below=1] {};
  	\node[phantom node] (3) [right of=1,above=1] {};

  	\path[every node]
   	(1) edge[loop above] node [above] {no$^{1}$} (1)
    	(1) edge[bend right] node [below] {no$^{2}$} (2)
    	(1) edge[bend left] node [above] {yes} (3);
	\end{tikzpicture}
\begin{tabular}{|l|}
\hline
    $^{1}$ $\mid${\it args}$\mid$ > 1 \\
\hline
    $^{2}$ $\mid${\it args}$\mid$ = 1 \\ 
\hline
\end{tabular}
\\
    \hline
    \end{tabular}
    \caption{Contrasting state machine architecture fragments for disambiguation, using individual states for each object (L) and a single state (R) where transitions are also based on conditions on the set of available arguments for disambiguation (1, 2) at the time the agent enters the disambiguation state.}
    \label{fig:archs}
\end{figure}
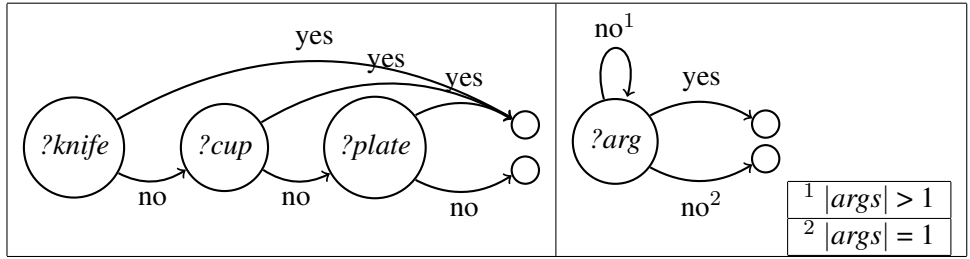

Evaluating the transition relation against conditions on the arguments being evaluated over at runtime necessitates storing these arguments as symbols elsewhere.  In our implementation we elected to use a pushdown automaton (PDA), as the context-free grammar of the interaction is Turing equivalent to a nondeterministic PDA, and to disallow operations on stack symbols other than the topmost.  Therefore, we can store existing conversational context, such as the object that is the current focus of conversation, on the stack symbol of a PDA.  In the disambiguation example, a ``no" input pops the stack to the next option, while a ``yes" answer may rewrite or push a new stack symbol.  The stack symbol can be constructed to store whatever information is necessary for the interaction.  Our implementation typically stores indicated objects and regions, objects that are being grasped by the agent, and options for objects and actions needing disambiguation.

We implement some modifications to the traditional structure of a PDA, which come from the requirements for using situatedness to establish context and composing information in real time.  As noted in Fig.~\ref{fig:archs}R, the two ``no" transitions shown differ not in the particular value of the argument being evaluated in the disambiguation state, but in the {\it conditions} on the set of possible arguments when entering that state.  This is important given the continuous nature of the world.  Using a PDA instead of an FSA is of no help in reducing the number of states in the search space if multiple transitions from the same state on the same input symbol still have to be generated for every particular value of a continuous stack symbol parameter.  For example, a coordinate indicated by deixis can move continuously through the 3D world, particularly with noisy vision and interpolated gesture recognition, and it is much more useful and concise to create a condition in the transition relation to check if the indicated coordinate is ``not null" or falls within a certain area than to, e.g., create one transition if the indicated coordinate is $(0.0,2.7,-0.4)$ and another if it is $(0.1,2.7,-0.4)$.

We add two operations to the traditional PDA set of {\sc Push}, {\sc Pop}, and {\sc Rewrite}.  The agent may need to disregard some or all preceding context, e.g., on the aborting or completion of an action, so {\sc Flush} clears the stack, and the stack symbol, except for information that persists physically, such as objects held by the agent.  {\sc PopUntil} functions similarly, but takes a state as a content argument, and pops the stack until the stack symbol equals what it was in the previous occurrence of that state (this is equivalent to {\sc Flush} if the specified state has never been entered previously).

\section{Situatedness, Composition, and Reasoning}
\label{sec:scr}
Now imagine that in prior to entering the disambiguation loop in Fig.~\ref{fig:archs}, the human has already specified an action to be undertaken with whatever object is to be singled out via the available disambiguation strategies.  Once the object has been successfully indicated, the action, which may have been defined many states ago, must be retrieved and applied to the object.  In continuation-passing style (CPS), this information is specified as the ``what to do next" argument \citep{van2010computational} as in a Montague grammar \citep{barker2004continuations}, and can be represented using \citeauthor{van2010computational}'s CPS function-application over the action denoted and object emerging from the disambiguation loop, as shown in a Haskell fragment:

{\tt cpsApply :: Comp (a -> b) r -> Comp a r -> Comp b r}

{\tt cpsApply m n = $\textbackslash$ k -> n ($\textbackslash$ b -> m ($\textbackslash$ a -> k (a b)))} \\

{\tt intAct\_CPS :: WorldState -> Action -> Comp (Object -> Bool) Bool}

{\tt intAct\_CPS bs (Action act obj) = cpsApply (intTAct\_CPS bs act)} \\ \hspace*{12mm} {\tt (intObj\_CPS obj)} \\

%optional
\iffalse
{\tt intTAct\_CPS :: WorldState -> Gesture -> Comp (Loc -> Loc -> Bool) Bool}

{\tt intTAct\_CPS bs Move = cpsConstAct move bs} \\

{\tt cpsConstAct :: (WorldState -> a) -> WorldState -> Comp a r}

{\tt cpsConstAct c bs = \k -> k (c bs)} \\

{\tt cpsConst :: a -> Comp a r}

{\tt cpsConst c = \ k -> k c} \\

{\tt cpsConstAct :: (WorldState -> a) -> WorldState -> Comp a r}

{\tt cpsConstAct c bs = \k -> k (c bs)}
\fi
%end

For example, we can specify a method to execute at the moment of state transition that will retrieve the action specified, apply it to objects or locations indicated by deixis, and prompt the agent to ask appropriate questions using possible interpretations of the action+object/location composition and present them to its interlocutor. 

\begin{wrapfigure}{r}{.5\textwidth}
\vspace{-6mm}
\begin{tikzpicture}[->,auto,node distance=3cm,
  thick,main node/.style={circle,draw},phantom node/.style={circle,draw}]

  \node[main node] (1) {{\it InterpDeixis}};
  \node[main node] (2) [right of=1] {{\it DisambTarget}};
  \node[phantom node] (3) [right of=2] {};

  \path[every node]
    (1) edge[bend left] node [above] {null, $\mathcal{A}$} (2)
    (2) edge[loop above, distance=1cm] node [above] {no, $\mathcal{B}$} (2)
    (2) edge[loop below, distance=1cm] node [below] {no, $\mathcal{C}$} (2)
    (2) edge[bend left] node [above] {yes, $\mathcal{C}$} (3);
\end{tikzpicture} \\
\begin{tabular}{|l|l|}
\hline
    $\mathcal{A}$ & [$o_1$, $o_2$, $(x,y,z)$], $\lambda w$.put($b$,on($w$))@$o_1$ \\
\hline
    $\mathcal{B}$ & [$o_2$, $(x,y,z)$], $\lambda w$.put($b$,on($w$))@$o_2$ \\ 
\hline
    $\mathcal{C}$ & [$(x,y,z)$], $\lambda w$.put($b$,on($w$))@$(x,y,z)$ \\
\hline
\end{tabular}
\vspace{-4mm}
\caption{\label{fig:PDA-CPS}PDA disambiguation fragment with continuation-passing style and function application on stack symbol.}
\vspace{-6mm}
\end{wrapfigure}

Fig.~\ref{fig:PDA-CPS} shows the steps through a deictic interpretation and disambiguation step with a previously-specified ``put" action on the stack with no destination yet specified.  The transition from {\it InterpDexis} to {\it DisambTarget} executes a function that supplies three possible destinations to stack symbol $\mathcal{A}$: objects $o_1$ and $o_2$ and location $(x,y,z)$.  The subsequent stack symbols $\mathcal{B}$ and $\mathcal{C}$ are created by {\sc Pop}ping in the ``no" transitions, returning to the same state.  At each step, the next destination option is applied to the variable $w$ in the action predicate until the human confirms that $(x,y,z)$ is the desired destination.  By exploiting continuation-passing style we can raise the type of the objects or location to the type required by the action ``put."

Our HAI system is also capable of one-shot learning for iconic gestures to indicate grasping particular objects, such as learning that miming holding a cup (such as part of the American Sign Language sign for ``cup") is an instruction to grasp the cup in that pose.  Having learned this instruction, the human can instruct the avatar to grasp the cup with a single gesture instead of indicating the cup first and then the grasping action.  However, this can also be used to fill in gaps in the existing context as part of action sequences other than ``grasp."  The VoxML encoding for $put(x,y)$ contains a $grasp(x)$ precondition before the actual object movement.  Thus, if the agent enters a state where the stack symbol contains an action with an outstanding variable---$\lambda b$.put($b$,$v$)---and the human supplies the iconic gesture for $grasp(cup)$, the avatar can directly lift the type $e \rightarrow t$ from $grasp(cup)$ to $\lambda b$.put($b$,$v$) and apply the argument $cup$ to $b$: $\lambda b$.put($b$,$v$)@$cup$ $\Rightarrow$ put($cup$,$v$).

\iffalse
Why PDA structure?
The vocabulary of acceptable inputs can be defined as a context-free grammar:

        // define the grammar
		/*
		// O: define the object
		// A: define the action
		// D: disambiguate
		// d: deictic gesture
		// V: verb phrase
		// N: noun phrase
		// y: positive acknowledgment (S,G)
		// n: negative acknowledgment (S,G)
		// i: static iconic gesture (indicates object)
		// a: dynamic iconic gesture (indicates action)
		S ::= OA|AO
			O ::= d|dD|v|vD|c|cD|s|sD
			A ::= a|aD|v|vD
			D ::= d|dD|v|vD|c|cD|s|sD|y|yD|n|nD
			*/

		// input symbols: received messages
		// stack symbols: array of state variables

The CFG can be rewritten as RTN or FSA
Multiple objects in situation (i.e., multiple candidates for reference in context) means we need a way to store this contextual information if we don't want to have new states for every possible contextual configuration
Contextual information can instead be stored as the stack symbol of a PDA
A world of continuous/non-discrete space requires that the stack symbol be evaluated relative to conditions instead of particular values
Speech acts (nevermind) and shifts in dialogue focus introduce flush and popuntil
\fi

\section{Discussion and Conclusions}
\label{sec:discussion}

The nondeterministic PDA architecture presented facilitates multimodal reasoning and interaction in real time.  Implementationally, we exploit the continuation-passing style available in the C\# language to use it with the Unity game engine on which VoxSim is built.\footnote{VoxSim repository at https://github.com/VoxML/VoxSim}  There may be cases where simpler or more restrictive behaviors are needed, while still requiring access to the contextual information provided by the agent's situatedness relative to the human and the world.  In these cases, the nondeterministic PDA serves as a general case of a deterministic PDA (where probabilities on all transition arcs equal 1), a nondeterministic finite automaton (where the stack symbol is always {\sc null}), or a standard deterministic FSA (where all probabilities are 1 and the stack symbol is {\sc null}).

Continuation-passing style as a method of incrementally aggregating contextual information through a discourse functions with all these methods.  Methods of any return type can be executed in state transitions as long as the return type can be raised to the type required by the calling function, and this makes it effective at composing from multiple modalities in real time.

% \newpage
 
\begin{acknowledgements} 
\noindent
This work is supported by a contract with the US Defense Advanced Research Projects Agency (DARPA), Contract CwC-W911NF-15-C-0238.  Approved for Public Release, Distribution Unlimited. The views expressed are those of the authors and do not reflect the official policy or position of the Department of Defense or the U.S. Government.  We would like to thank Kelley Lynch for the Haskell implementation code snippet, and Kyeongmin Rim for work on the VoxSim platform. 
\end{acknowledgements} 

\vspace{-0.25in}

{\parindent -10pt\leftskip 10pt\noindent
\bibliographystyle{cogsysapa}
\bibliography{References}

}

% Leave a blank line before the closing brace to ensure the final 
% reference has the proper indentation. 

\end{document}